# openACC. An open database of car-following experiments to study the properties of commercial ACC systems.

Michail Makridis[1,b,a], Konstantinos Mattas[a], Aikaterini Anesiadou and Biagio Ciuffo

[a] European Commission – Joint Research Centre., Ispra (VA), Italy
[b] ETH Zürich, Institute for Transport Planning and Systems (IVT). Zürich, CH

**ABSTRACT**

Commercial Adaptive Cruise Control (ACC) systems are increasingly available as standard options in modern vehicles. Their penetration rate in the fleet is constantly increasing, as well as their use, especially in freeway conditions. At the same time, still little information is openly available on how these systems actually operate and how different is their behavior, depending on the vehicle manufacturer or model. This represents an important gap because as the number of ACC vehicles on the road increases, traffic dynamics on freeways may change accordingly, and new collective phenomena, which are only marginally known at present, could emerge. Yet, as ACC systems are introduced as comfort options and their operation is entirely under the responsibility of the driver, vehicle manufacturers do not have to provide any evidence about their performances and thus, any safety implication connected to their interactions with other road users escapes any monitoring and opportunity of improvement.

To reduce this gap, the present paper summarizes the main features of the openACC, an open-access database of different car-following experiments involving a total of 16 vehicles, 11 of which equipped with state-of-the-art commercial ACC systems. As more test campaigns will be carried out by the authors, OpenACC will evolve accordingly. The activity is performed within the framework of the openData policy of the European Commission Joint Research Centre with the objective to engage the whole scientific community towards a better understanding of the properties of ACC vehicles in view of anticipating their possible impacts on traffic flow and prevent possible problems connected to their widespread. A first preliminary analysis on the properties of the 11 ACC systems is conducted in order to showcase the different research topics that can be studied within this open science initiative. Interesting dimensions for research include the acceleration and deceleration distributions of ACC systems under normal car-following conditions, their response time to perturbations, their desired time headway settings, their string stability, as well as other phenomena such as their tractive energy demand, driving behavior and homogeneity, and their effect on traffic safety and network capacity. For comparison purposes, results also include measurements from the vehicles taking part to the experiments and not using the ACC. The goal of the openACC database is to engage researchers on conducting studies in various topics using the same datasets, set the foundations for a large universal experimental dataset and help towards reducing the observed complexity in current transport networks. In this light, openACC, over time, also aims at becoming a reference point to study if and how the parameters of such systems need to be regulated, how homogeneously they behave, how new ACC car-following models should be designed for traffic microsimulation purposes and what are the key differences between ACC systems and human drivers.

---

[1] Corresponding author. Email: mmakridis@ethz.ch

# 1 INTRODUCTION

Advanced in-vehicle technologies promise to disrupt road transportation as it is known today bringing benefits in several dimensions. More specifically, vehicle automation and connectivity aim to assist towards reducing traffic congestion, facilitating traffic flows, minimizing travel times, pollutant emissions and improving traffic safety and comfort. The automotive industry has made significant investments on these technologies with the objective to compete on a potentially multi-billion market (European Commission, 2019). Even if full vehicle automation will still require several years to be achieved, partial levels of automation are already oavailable on the market. The Society of Automotive Engineers (SAE) has proposed a classification based on 5 vehicle automation levels (SAE International, 2018). Many existing vehicles are classified as level 1 or 2 as they are able, under the constant supervision of the driver, to automatically keep longitudinal and/or lateral control of the vehicle. The system in charge of maintaining the longitudinal vehicle control is called the Adaptive Cruise Control (ACC) and is one of the most widespread driving assistance systems available in the market.

More than 20 years have passed since the first introduction of ACC-equipped vehicles in the market (Xiao and Gao, 2010). Despite the research efforts, the operational design of these systems is still almost a black box to the research community, as it is usually protected by intellectual property rights. A review that summarizes the control strategies for ACC systems in experimental vehicles can be found in the work of He et al. (He et al., 2019). Moreover, the operational differences between commercially available ACC systems of different manufacturers is not yet thoroughly studied in the literature. For this reason, many assumptions are usually necessary in order to study the impact that mass deployment of ACC equipped vehicles will have on public networks. In fact, in the literature, the properties of ACC systems and their impact in terms of traffic flow, stability, emissions and traffic safety are mostly discussed through simulation studies (Hu et al., 2019; Li and Wagner, 2019; Liu et al., 2018; Makridis et al., 2020a; Mattas et al., 2018; Olia et al., 2018; Shladover et al., 2012; Sun et al., 2018; Wang et al., 2018; Ye and Yamamoto, 2018; Yu et al., 2018). However, simulation results are limited by a large number of known factors (data quality, models' design and accuracy, calibration process, limited data etc.).

On the other side, experimental campaigns are more reliable for drawing conclusions, but they demand more resources, they are usually constrained in a small region, and they are difficult to organize and correctly conduct. Recent technological advances in data acquisition systems and computing and the increasing interest in the impact of automation and connectivity has called for new experimental campaigns able to provide new elements to the scientific community. The experimental campaigns can be categorized in two main clusters; those who focus on the supervision of a section or area in order to capture complex traffic phenomena and; those who zoom in to the vehicle technology in order to understand better the properties of newly deployed advanced vehicle systems.

In the first category, NGSIM ("NGSIM," 2006), although developed 15 years ago, is still probably the most well-known database and it has been extensively used in traffic microsimulation research. NGSIM dataset is a collection of real-world trajectories, based on the use of cameras mounted on tall buildings and covering approximately a 1-km long roadway section with a frequency of 10Hz in several US locations. Recently, Barmpounakis and Geroliminis conducted the pNEUMA large-scale field experiment (Barmpounakis and Geroliminis, 2020) aiming to record traffic streams in a multi-modal congested environment over an urban area using Unmanned Aerial Systems. The dataset was generated by a swarm of 10 drones hovering over a traffic intensive area of 1.3km$^2$ in the city center of Athens, Greece, covering more than 100km-lanes of road network at 25Hz. This initiative allows the deep investigation of critical traffic phenomena in urban areas as never has been possible in the past. Additionally, Krajewski et al. (Krajewski et al., 2018) created the highD (highway drone dataset) with detailed vehicle trajectories over a road segment of around 420m, from a drone hovering next to German highways at 25Hz. All these datasets can be particularly useful for studying traffic related phenomena, but they don't provide technical specification for the vehicles involved in the tests and thus they cannot be used for studying the various in-vehicle technologies.

In the second category, there are several studies with small-scale car-following experiments focusing on understanding the in-vehicle technology and the driving behavior that this generates. Milanes and Shaldover (Milanés and Shladover, 2014) carried out experimental tests using up to four production Infiniti M56s provided by Nissan. The vehicles were equipped with a commercial ACC system that uses a LIDAR for detecting the preceding vehicle. The aim of this work was to compare the performance of three different longitudinal controllers, a production ACC, the IDM model and a Cooperative-ACC system. Furthermore, Knoop et al. (Knoop et al., 2019) conducted an experiment with seven SAE level-2 vehicles driven in a car-platoon formation. The vehicles were driven on public roads in the Netherlands for a trip of almost 500 km. The authors discuss the observed instability in the car-platoon when all vehicles have ACC activated. Severe variations in the speed were also observed, leading in cases to discomfort and even risks of rear-end collisions. However, since the experiment was not performed inside a controlled environment, occasionally, there was external interference from other network users. The data acquisition was performed using commercially available devices (Ublox) that have inferior accuracy in comparison to a differential GNSS. Furthermore, in Gunter et al. (Gunter et al., 2019), the authors assessed the string stability of seven 2018 model year ACC-equipped vehicles from two makes. All vehicles under all following settings were found to be string unstable. Ublox devices were used for the data acquisition in this case as well, focusing solely on doppler-based speed measurements. String stability of car-platoons on empirical observations is briefly discussed also in Makridis et al. (Makridis et al., 2020b). On another dimension, the energy impact of ACC on car-following conditions is discussed in (He et al., 2020a). Both studies are based on partial trajectories from the openACC database but both have a limited scope and provide only preliminary observations on the behavior of the ACC systems. Finally, Makridis et al. (Makridis et al., 2019c) shows through a car-following campaign with a commercially available ACC system that the reaction time of the controller is around 1.1s and the time gap has been found distinctly larger than that of a human driver. Only few of the above databases are available to the research community for further investigation. A common missing element from all the above studies is that they don't focus on the heterogeneity in the different commercially available designs of the ACC systems. In addition, since the data collected in these studies are not publicly available, the contribute to a better understanding of the implications of the ACC widespread only to the extent of the analysis carried out by the respective authors.

To overcome this limitation the authors of the present paper, in line with the open data policy of the European Commission have decided to make publicly available the complete database resulting from different experimental campaigns carried out in the past years and involving several commercial ACC systems. In this light, the present paper summarizes the main features of the openACC database (Joint Research Centre, European Commission, 2020), an open-access database of different car-following experiments involving 16 vehicles, 11 of which equipped with state-of-the-art commercial ACC systems. As more test-campaigns will be carried out and the related data extracted, openACC will also evolve and become more complete. The objective is to engage the whole scientific community towards a better understanding of the properties of ACC vehicles in view of anticipating their possible impacts on traffic flow, prevent possible problems connected to their widespread and define the functional requirements that future vehicles will need to have to avoid these problems.

At present the database includes the results of three different test campaigns by means of which the research team wanted to understand the behavior of commercial ACC systems. First, a small car-following campaign was organized on public freeways with two and three vehicles in car-platoon formation. The personal experience of discomfort on speed perturbations for the passengers of the following vehicles when the ACC was enabled was the motivation of a more challenging field test. A second bigger real-world campaign was thus organized with five vehicles in car-platoon formation. Data from the second experiment revealed very interesting findings on basic properties of the ACC controllers such as the response time and time headway and question the systems' ability to ensure string stability in the car-platoon. Finally, a third campaign was conducted in a proving ground with 5 high-end vehicles in order to provide insights on the above issues in a more systematic way with a more precise data acquisition system and study more thoroughly the observed string instability issues while avoiding a lot of the external influences that are present in the public roads. The trajectory datasets have a sampling rate of 10Hz. Preliminary discussion within this work on the datasets provides a first feedback to researchers, the industry and policy makers on why the construction of such a database is important to all stakeholders, what kind of conclusions can be made based on such tests and finally clarifies the added value of extending the openACC database with additional tests.

The paper is organized as follows. The next section provides a brief description of each field test and the main specifications of the vehicles involved in the experiments. Section 3 discusses the data post-processing procedure. Section 4 presents the preliminary results of all tests and finally the last section provides the summary of the main findings and future work on the topic.

## 2  Experimental campaigns

The first version of the openACC database includes the results of three distinct car-following campaigns. The first two test campaigns were carried out on public freeways and consisted of different experiments involving 2, 3 and 5 ACC equipped vehicles moving in car-platoons. Retrieving data from the first two test campaigns was challenging. As also reported by Knopp et al. (2019), indeed, car-following experiments on public motorways are usually challenged by the behavior of the surrounding vehicles and the road geometry and therefore, it is not always easy to preserve the platoon for very long distances. In addition, motorway driving usually limits the speed range that can be analyzed. The shortest time headway setting for each vehicle driven by the ACC system was used to avoid cut-in situations from other users. Furthermore, a post-processing stage had to be applied in order to fix the sampling rate at a constant frequency of 10Hz. For these reasons a third test campaign was conducted in the AstaZero[2] proving ground with five vehicles and a high-fidelity differential GNSS data acquisition system, in order to deal with the above-mentioned issues and provide a more complete study on the issues observed on public roads.

Figure 1 illustrates the path layout for the three campaigns whereas the basic specifications of the vehicles involved in the experiments are given in Table I. Vehicles with prefix "(L)" were used as leaders in the car-platoon tests. Finally Table II presents general information of the three campaigns.

---

[2] http://www.astazero.com/

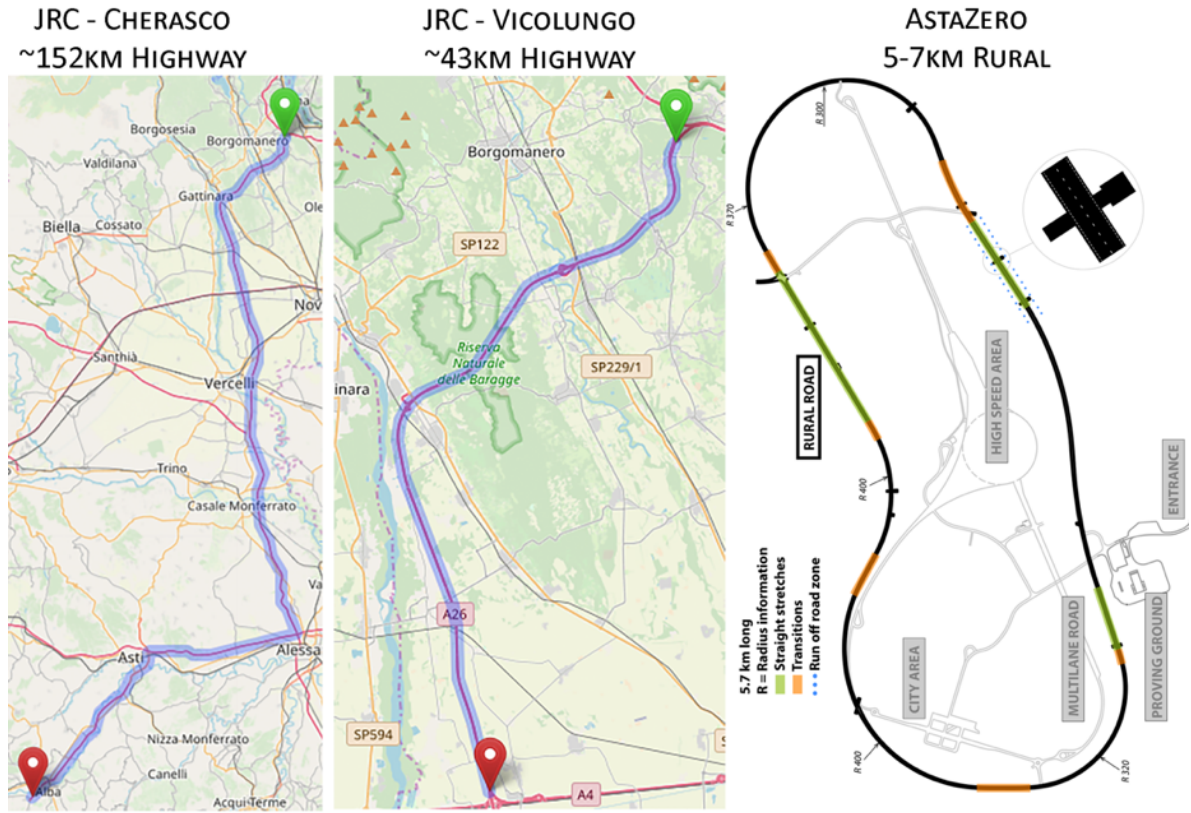

**Figure 1.** Path layout for the three test campaigns included in the first version of the openACC database.

**Table I.** The main specifications of the vehicles involved in the experiments.

| Vehicles | Max power (kW) | Drive-Fuel | Engine displacement (cc) | Battery capacity (kWh) | Propulsion type | Top speed (km/h) | Model year |
|---|---|---|---|---|---|---|---|
| (L) Fiat (500X) | 103 | diesel | 1956 | - | ICE | 190 | 2016 |
| Volvo (XC40) | 140 | diesel | 1969 | - | ICE | 210 | 2018 |
| (L) VW (Polo) | 63 | Gasoline and liquid propane gas | 1390 | - | ICE | 177 | 2010 |
| Hyundai (Ioniq hybrid) | 104 | gasoline | 1580 | 1.56 | HEV | 185 | 2018 |
| (L) Mitsubishi (SpaceStar) | 59 | gasoline | 1193 | - | ICE | 173 | 2018 |
| KIA (Niro) | 77.2 | gasoline | 1580 | 8.9 | PHEV | 172 | 2019 |
| Mitsubishi (Outlander PHEV) | 99 | gasoline | 2360 | 12 | PHEV | 170 | 2018 |
| Peugeot (5008 GT Line) | 130 | diesel | 1997 | - | ICE | 208 | 2018 |
| VW (Golf E) | 100 | electricity | - | 35.8 | BEV | 150 | 2018 |
| Mini (Cooper) | 100 | gasoline | 1499 | - | ICE | 210 | 2018 |
| Ford (S-Max) | 110 | diesel | 1997 | - | ICE | 196 | 2018 |
| (L) Audi (A8) | 210 | diesel | 2967 | - | ICE | 250 | 2018 |
| Tesla (Model 3) | 150 | electricity | - | 79 | BEV | 210 | 2019 |
| BMW (X5) | 195 | diesel | 2993 | - | ICE | 230 | 2018 |
| Mercedes (A Class) | 165 | gasoline | 1991 | - | ICE | 250 | 2019 |
| Audi (A6) | 150 | diesel | 1968 | - | ICE | 246 | 2018 |

**Table II.** Overview of the first version of the openACC dataset.

| Campaign | Ispra-Cherasco (N.1) | Ispra-Vicolungo (N.2) | AstaZero (N.3) |
|---|---|---|---|
| **Leading vehicles** | Fiat 500X, Volvo XC40, VW Polo | Mitsubishi SpaceStar | Audi A8 |
| **Driving modes of leading vehicles** | Human | Human | ACC |
| **Vehicles involved in car-following** | Hyundai Ioniq hybrid, Volvo XC40 | KIA Niro, Mitsubishi Outlander PHEV, Ford S-Max, Peugeot 3008 GT Line, VW Golf E, Mini Cooper | Tesla Model 3, BMW X5, Mercedes A Class, Audi A6 |
| **Driving modes of following vehicles** | Human, ACC | Human, ACC | Human, ACC |
| **Driving duration in openACC (h)** | 4.99 | 6.3 | 12.6 |
| **Driving distance in openACC (km)** | 493 | 641 | 833km |
| **Frequency** | 10 Hz | 10 Hz | 10 Hz |

**Experimental Campaign N.1 – Driving from Ispra (VA) to Cherasco (CO)**

The first campaign was conducted in the last quarter of 2018 and involved two days of car-following testing, either with two or with three vehicles in car-platoon formation on the public freeway roads from Ispra (VA) to Cherasco (CO) in northern Italy. The experiments were scheduled to take place after peak morning hours and during lunch time hours, aiming to avoid the disturbances from other road users. All vehicles were equipped with Ublox 8 GNSS data acquisition devices. The Antenna-to-Front-Bumper and Antenna-to-Rear-Bumper distances were measured for accurate computation of the intervehicle spacings. The resulting sampling rates after interpolation were around 10Hz.

The goal for the leader was to drive manually on free-flow, without following other road users. The leader was instructed to drive with the car manually and to create occasionally small perturbations in a random yet realistic way through deceleration and acceleration around the desired speed. The follower was instructed to drive whenever possible with the ACC enabled.

**Experimental Campaign N.2 – Driving from Ispra (VA) to Vicolungo (NO)**

This campaign was conducted in the first quarter of 2019 and involves three days of car-following testing from Ispra (VA) to Vicolungo (NO) and back, in northern Italy. The testing was performed with five vehicles of various brands and models driving on car-platoon formation. As in the first campaign, the tests were scheduled for non-peak hours in order to minimize the disturbances from other road users. The same data acquisition system with campaign N.1 was used.

This campaign was a follow-up of the first one. In the first campaign there were signs about unstable behavior of the ACC controller. In some perturbations, the driver of the following vehicle experienced a feeling of discomfort due to the lag in the response of the ACC. Therefore, the second campaign was organized in order to provide more insights on this issue due to the longer car-platoon. The leader was instructed to drive manually and perform occasional random deceleration and accelerations over a desired speed in a realistic way. The followers, whenever possible were driving with ACC on.

Fig. 2 depicts some of the challenges encountered during the real-world experiment, showing four indicative examples with physical and artificial obstacles encountered during the campaign. More specifically, Fig. 2 a) shows a cut-in situation when a road user needed to drive on an off-ramp, Fig. 2 b) is an example when the car-platoon consistency brakes on tolls, Fig. 2 c) is another example inside a tunnel when the GNSS receiver provides position data of low quality and Fig. 3 d) shows a situation when a slow moving truck was an obstacle for the experiment and overtaking had to be performed. Manual inspection of the dataset was performed in order to remove visually visible noisy trajectory parts and improve the quality of the car-following datasets. The implementation of a more sophisticated filtering workflow, able to detect and isolate the above-mentioned challenges is currently in ongoing process.

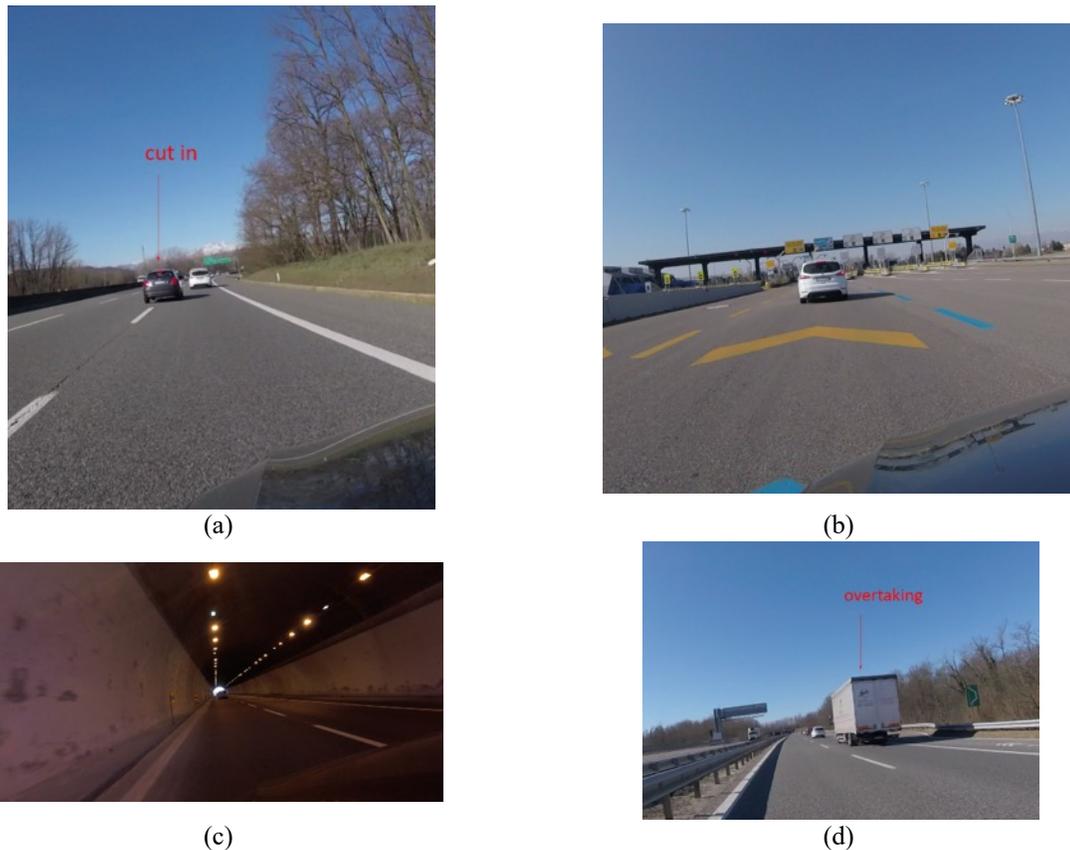

**Figure 2.** Challenges encountered during the experiment, Fig. a) is a cut-in situation due to an off-ramp, Fig. b) is car-platoon inconsistency due to tolls, Fig. c) is a situation with low-quality GNSS signal due to a tunnel and Fig. d) is a lane change and overtaking situation due to a slow truck ahead.

### Experimental Campaign N.3 – AstaZero proving ground

This last campaign was conducted in the second quarter of 2019 and it involves two days of car-following testing on the Rural road of the AstaZero test track in Sweden. The Rural road is approximately 5.7 km long. The campaign involves five high-end vehicles, from four different makes, all different models. The system used for trajectory data acquisition is the RT-Range S multiple target ADAS measurements solution by Oxford Technical Solutions Company, with a differential GNSS accuracy.

Having significant insights from the first two campaigns about high response times of the ACC controllers, large headway settings and instability, this campaign was organized in a way to become a research tool for more systematic study of the ACC behavior and properties. Moreover, it took place within a protected environment, in order to avoid risks and limitations of tests on public roads. The experiments are organized in laps, where one lap corresponds to the path shown in Fig. 1. In all the tests, the leading vehicle is the same and it is driven with the ACC enabled in order to avoid noisy fluctuations around the desired speed due to manual maneuver. In general, two different car-following patterns are applied for the following vehicles, a) car-platoon with constant speed and b) car-platoon with perturbations of the target speed (deceleration to a new desired speed) from an equilibrium point. For the second pattern, a radio-based communication between the drivers of the first and last vehicles ensures that the speed of the last vehicle is stable on the desired speed and therefore the car-platoon is close to an equilibrium state, prior to applying a new perturbation. When the driver in the leading vehicle wants to apply a new perturbation, he sets the desired speed of the ACC system to a new lower desired speed value, the vehicle decelerates using the ACC system and then the driver resets the desired speed to the previous setting. The duration of the perturbation is automatically adjusted based on the deceleration strategy applied by the controller. This procedure is selected to perform the different perturbations in a controlled way, and it resembles the way that vehicles with ACC enabled behave on road. For safety reasons in each lap the desired speeds were fixed to 13.9 m/s – 16.7 m/s along the curves and to 25 m/s-27.8 m/s on the straight parts. The differential GNSS system used for the data acquisition ensures precision of 2 cm/s in the speed and 2 cm in the positioning measurements with a sampling rate around 100Hz, which is down sampled to 10Hz as described in the Section 3.

# 3 Data post-processing

Trajectory and speed vehicle data were acquired with two different systems for the three experimental campaigns. Regarding the on-road experiments, data acquisition was obtained in binary format from the U-blox M8 devices, one device installed per vehicle. U-blox M8 devices include accelerometers, gyroscopes and GNSS receivers. The average horizontal accuracy reported by the receivers is less than 50 cm. Regarding the test track experiment, data acquisition was performed with OXTS inertial navigation system. Due to the different data acquisition methods both in terms of accuracy and systems, data post-processing procedures are different. It is worth underlining here that the post-processing procedure does not imply any correction of the raw data (e.g. data filtering or smoothing) but only the identification of specific circumstances in which the data can be enriched or for which the data should be removed in order to simplify the use of the data themselves and to avoid that by using them misleading conclusions can be achieved. The following post-processing activities have been indeed carried out:

- Regarding the real-world data, in some measurements and for some vehicles it was possible to access vehicle OBD or CAN data. Such type of data is incomplete and in most cases, it does not provide enough accuracy for trajectory/speed series reconstruction. However, they provide the acceleration pedal positions or driving assistance information, and whenever available during the campaign, they were used to auto-detect when the ACC of the vehicles in the platoon is activated. This information is available to the users in the public openACC dataset.
- Visual inspection of speed series was performed in order to detect parts in the speeds and trajectories with outliers or noise. These parts have been removed and are not included in the public openACC dataset.
- On average U-blox M8 data, which was used on real-world campaigns, provide measuring frequency of around 3-5 Hz depending on the GNSS reception. Data interpolation was applied on the first and second campaign in order to ensure a sampling frequency of 10 Hz. A piecewise cubic polynomial being twice continuously differentiable on subintervals of two points was used to interpolate the data. For the third campaign, the sampling rate derived by the OXTS system was more than 10 Hz (around 100Hz) and it was down-sampled to 10 Hz.
- The position of the antennas is translated to the front bumper of each vehicle. First, the instantaneous inter-vehicle distances (IVS) are calculated based on position data. Then these measurements are corrected based on computing bumper to bumper distances, by subtracting the leader's antenna-back bumper distance and followers antenna-front bumper distance.

The final dataset that is publicly available online contains three folders, one per experimental campaign. Each folder has a list of .csv files that correspond car-platoon data series. Each .csv file contains a series of columns as shown in the Table III. The folders contain a file with the specifications of the vehicles and a brief description with the experimental design. For illustrative purposes one of the files is reported as an annex to the present document.

**Table III** Denoting the columns contained in the .csv files of the database.

| Column ID | Description | Units |
|---|---|---|
| **Time** | Common time frame for all vehicles | $s$ |
| **Speed** | Raw Speed (Doppler) | $m/s$ |
| **Lat** | Latitude | $rad$ |
| **Lon** | Longitude | $rad$ |
| **Alt** | Altitude | $m$ |
| E | East (x) coordinate in the local ENU plane (common center for all vehicles) | – |
| N | North (y) coordinate in the local ENU plane (common center for all vehicles) | – |
| U | Up (z) coordinate in the local ENU plane (common center for all vehicles) | – |
| **VE** | Speed in the East direction of the local ENU plane | $m/s$ |
| **VN** | Speed in the North direction of the local ENU plane | $m/s$ |
| **VU** | Speed in the Up direction of the local ENU plane | $m/s$ |
| **IVS** | IVS computed from raw GNSS data after bumper to bumper correction. | $m$ |
| **Driver** | The driver of the vehicle: "Human" for manual driving, "ACC" for ACC driving. | – |

# 4  Preliminary Results

This open science dataset creates unique opportunities for researchers in order to observe commercially available automation controllers under real world conditions. The aim of this section is to present some preliminary results from each dataset and give few concrete examples on the topics where researchers can work on, some of which are outside of the authors' research activities. This offers the opportunity to understand better in-vehicle technologies, explain (partially) the vehicle's on-road behavior and facilitate the development of more precise tools and models on the simulation domain. Additionally, openACC offers the possibility to perform comparison between ACC-driven and Human driven behaviors (although linked to the behavior of the specific drivers involved in the study). Finally, openACC creates the opportunity for research studies on the heterogeneity of the different ACC systems, give feedback to policy makers on the regulation of controllers, generate a discussion for traffic modelers on the simulation side of ACC-driven and automated vehicles and provide insights on how high penetration rates of such systems could impact traffic flow in future road networks.

The indicative results presented here focus on the following main aspects:

- Understanding the controllers' high-level behavior demonstrating their acceleration and deceleration operational domain under car-following conditions.
- Studying heterogeneous behavior of the ACC controllers from different manufacturers.

- Showing the basic properties of commercial ACC systems such as response time, time headway settings and speed/acceleration variations.
- Observing the string instability of multi-manufacturer commercial ACC systems while they move in car-platoon formation.
- Facilitate research on the traffic safety domain by studying the inter-vehicle spacings in congested (Jiang et al., 2015) conditions.
- Open a discussion on the traction energy demand of ACC systems, as well as the similarities and differences with human drivers.
- Provide insights on the effect that commercial ACC systems can have on road capacity.

To the best of the authors' knowledge, this is the first study that demonstrates the behavior of such heterogeneous and detailed car dataset on car-following conditions and at the same time demonstrates the behavior of commercially-available ACC systems regarding some of their most important properties. The resulting openACC uses a simple data format that can be provided by most data acquisition systems equipped with a GNSS and an antenna and it will be updated by the Joint Research Centre of the European Commission, adding wherever possible new car-following tests. External parties willing to include their ACC experiments to the openACC database are suggested to get in contact with the authors.

## 4.1 Vehicle dynamics on on-road tests

Commercial data acquisition devices offer good accuracy on capturing position and ground speed data especially on open spaces such as freeways and when the weather conditions are good. Retrieved data have enough accuracy for studying some of the topics discussed in this study such as the response time and time headway settings of ACC, string stability, driving behavior etc. However, differentiation in order to retrieve vehicle acceleration or jerks can be quite challenging and therefore further processing is needed for relevant studies such as computation of instantaneous energy consumption or jerk-related safety procedures. Figure 4 illustrates the acceleration and deceleration distributions from the first campaign as derived from raw data and after the application of a simple moving average filter, similarly to what is described in (Jiang et al., 2015), which produces more realistic acceleration values. The prefix 'L' before the name of the vehicle in the x-axis signifies that this vehicle was used as a leading vehicle at some point during the campaign. Unfortunately, only partial data of human-driven car-platoon are available in this test. The acceleration distributions of both following vehicles have much higher peak values with the ACC enabled than when the vehicle is manually driven. Additionally, it is interesting to observe that the distributions when the vehicles are ACC-driven have high outliers values. Intuitively, this was experienced also on the road during the tests when the leader performed perturbations and the controller reacted with a significant lag, which led to uncomfortably high deceleration values and fast approaching of the leader. This was never the case when the follower was driving in manual mode.

Fig. 5 a) refers to a part of campaign N.1, where the two followers drive with ACC on, while in Fig. 5 b) the following vehicle has no ACC, based on raw measurements. Before discussing the findings here, let's describe first what is depicted in each figure. Each figure consists of four subfigures. The top-most one is the trajectory plot of the vehicles. The second top-most one refers to the oblique trajectories (Cassidy and Windover, 1995; Munoz and Daganzo, 2002) that provide a better snapshot on the traffic dynamics. Oblique plots are a powerful alternative to visualize traffic conditions as it successfully tackles the scaling problem that appears in trajectory plots. For more information, we refer the reader to the above publications. The third and fourth subfigures show the corresponding speed and time headway profiles of the vehicles.

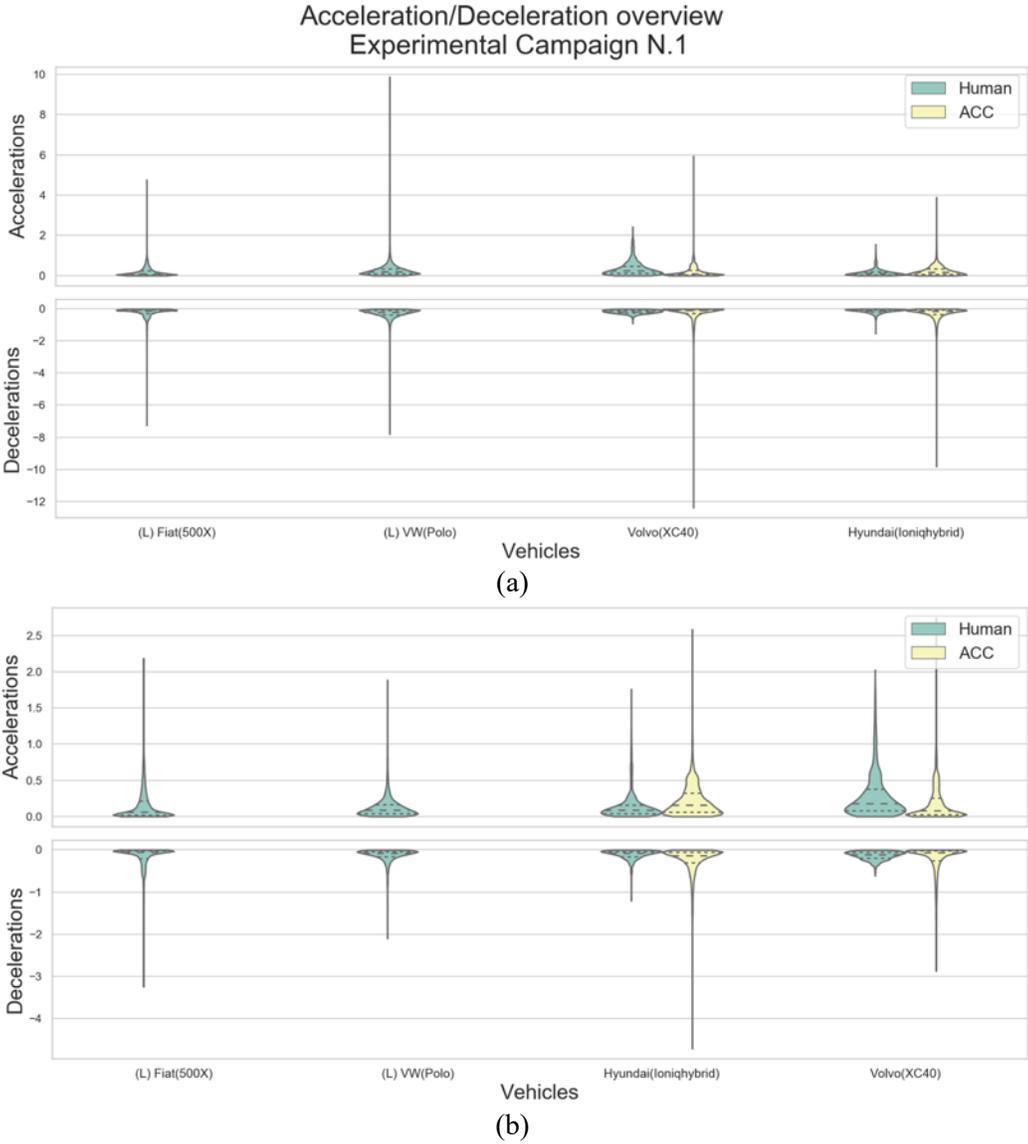

**Figure 4.** Acceleration and deceleration distributions per vehicle and driving mode. Prefix "(L)" before the name of a vehicle signifies that this vehicle was used as a leader of the car-platoon. Fig. 4a) depicts the raw measurements, while Fig. 4b) illustrates the same data after the application of a moving average filter.

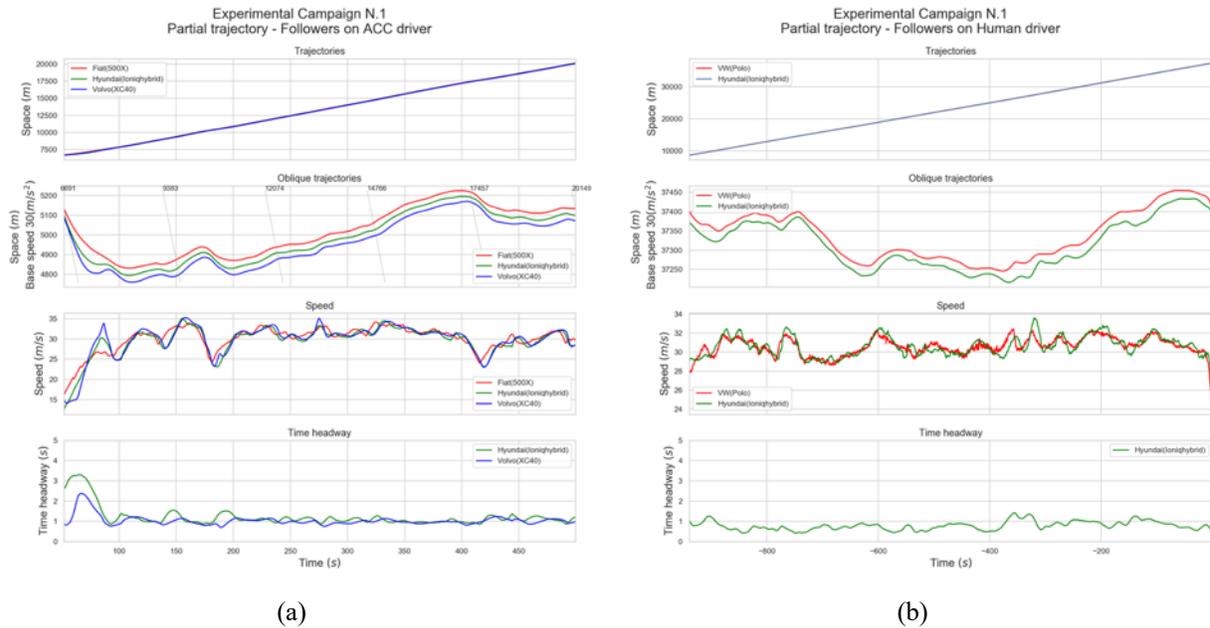

(a)                                      (b)
**Figure 5.** Two snapshots from campaign's N.1 trajectories. In Fig. 5 a) the two followers are driving with ACC on, while in Fig. 5 b) the following vehicle is human driven.

In Fig. 5 a) the leading vehicle is human-driven, while the speeds of the two followers are regulated by their ACC systems. The leader (red line) accelerates from 15m/s to 30m/s and then keeps an almost constant speed. The trajectories plots do not reveal the full dynamics. In the oblique plots, it is clear that there is a variance in the inter-vehicle spacings due to the acceleration of the leader. In terms of speeds, this creates an overshoot for the following vehicles, which is visible in the third subfigure. Finally, the headway series (fourth subfigure) show that the instantaneous headway increases a lot (reaching over 3s values) during the acceleration phase and then stabilizes around the headway setting value selected by the driver. These variations on the space and time domains have a direct impact on the capacity of the network and suggest more in-depth investigation.

Fig. 5 b) shows another trajectory snapshot with two vehicles without using the ACC. Here, there is an obvious variability on the speed domain, which is expected, since human drivers do not have a clear goal to keep a constant headway. An assumption could be that the distance from the leading vehicle is regulated by the driver's perceived level of safety, which in turn, it is based on the individual driver's characteristics and behavior. This is something that has been brought to attention in the literature (Jiang et al., 2015; Laval et al., 2014; Ngoduy et al., 2019), and this is the reason why an increasing number of authors suggest the need for more accurate reproduction of vehicle dynamics and driving behaviors in traffic modeling (Ciuffo et al., 2018; Fadhloun and Rakha, 2019; He et al., 2020b; Makridis et al., 2019a). In the whole dataset of the first campaign, the observable instantaneous time headway distributions for the shortest settings of the two ACC controllers have peak values around 0.9s and 1.3s for the Hyundai and the Volvo respectively.

The acceleration and deceleration distributions for the second campaign (N.2) per vehicle are shown in Fig. 6 based on raw data after the application of a moving average filter in order to provide realistic acceleration values. The visual representation shows similar behavior with the first campaign, as the ACC systems have higher peaks which after inspection of the dataset, they correspond to trajectory parts where the leading vehicle performed some perturbation. Looking at median values for the same vehicle (dashed lines in the violin plots) with and without ACC turned on, it seems that under stable conditions, which was the situation for the majority of the experiment, the ACC produces lower accelerations than human drivers. An assumption, which is also highlighted in the literature is that such controllers have been designed with a clear goal towards passengers' comfort. However, during the perturbations of the leading vehicles, a set of sharp acceleration and deceleration values appear. It is interesting to note that the commercial ACC systems by design can see one or in some only cases two vehicles ahead. Attentive human drivers can see many more vehicles ahead under congestion and thus they can better anticipate downstream perturbations much earlier. It should be mentioned that Mini Cooper whenever used was the last vehicle in the platoon and this justifies the higher peak acceleration and deceleration values in comparison with the rest of manually-driven vehicles. Even in this case however, the peak deceleration value of the Mini Cooper is lower than the corresponding values of the ACC systems that were upstream in the car-platoon formation.

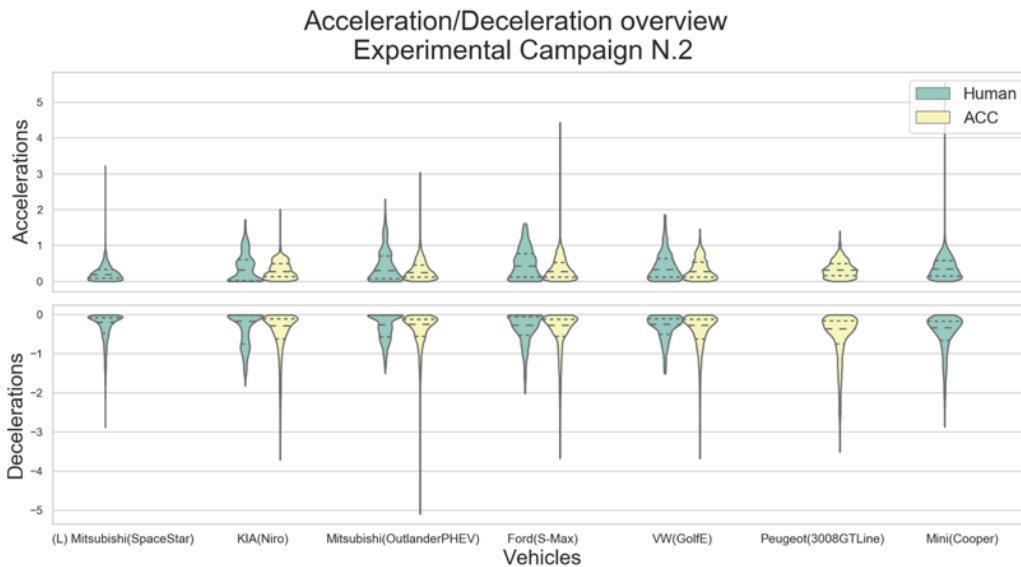

**Figure 6.** Acceleration and deceleration distributions per vehicle and driving mode. Label "(L)" before the name of a vehicle indicates that this vehicle was used as a leader of the car-platoon.

Fig. 7 shows two trajectory snapshots from campaign N.2 similar to Fig. 5. In Fig. 7 a) all vehicles drive with ACC on, while in Fig. 7 b) all vehicles have ACC off. In a car-platoon of five vehicles the insights from the first campaign are amplified and more concrete conclusions can be extracted. Human drivers do not respect the 'constant time headway' policy of the ACC controllers. The higher variation in the speeds of the manually-driven vehicles is obvious in comparison with the one from the ACC controllers.

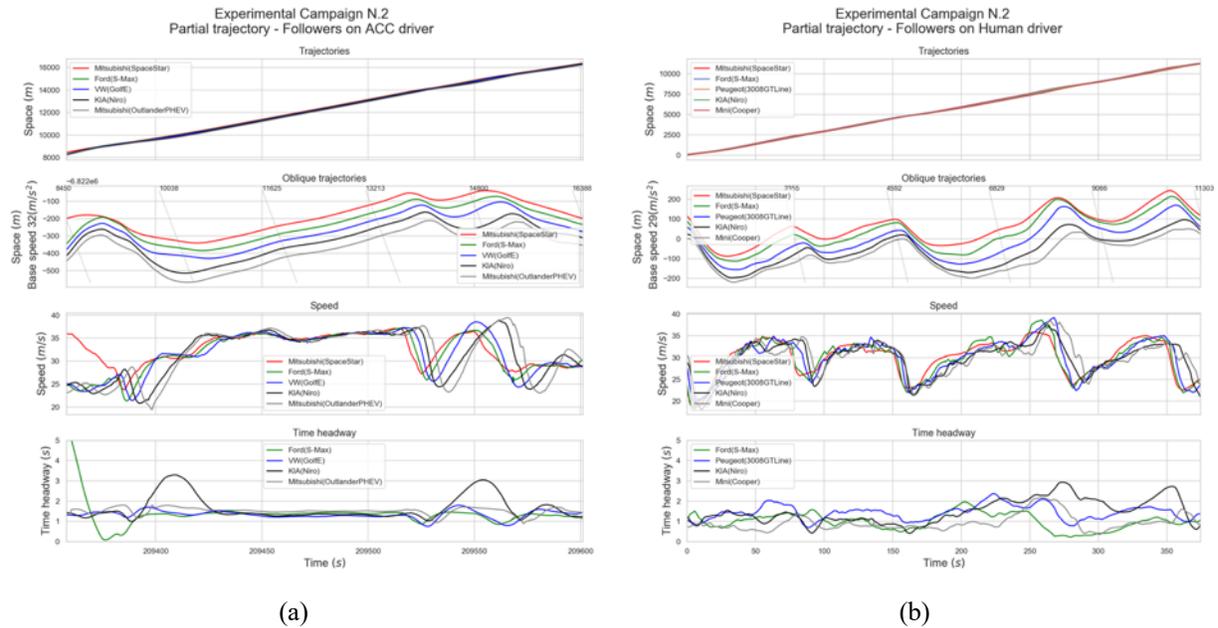

(a) (b)
**Figure 7.** Two snapshots from campaign's N.2 trajectories. In Fig. 7 a) the two followers are driving with ACC on, while in Fig. 7 b) the following vehicle is human driven.

Looking at the oblique plots, it is clear that the ACC vehicles on constant speeds keep constant gaps and thus they have almost equal time headways. The corresponding values for human-driven vehicles are quite random and as mentioned earlier they depend a lot on the driver's behavior and style. The most interesting finding in this campaign is related to string (in)stability. Looking at the third subplot of Fig. 7 a), it is apparent that the perturbation from the leader of the car-platoon is amplified upstream by the ACC controllers. The findings in Fig. 7 b) for trajectories with human-driven vehicles are quite different. Manual drivers anticipate the changes in the speed of the car-platoon leader and absorb the perturbation that occurs downstream. In terms of time headway values, there is no apparent pattern for all the vehicles involved in the test and thus it can be assumed that the inter-vehicle spacing derives based on the individual behavior of each driver.

Driver heterogeneity on car-following is reflected on time headway measurements (Saifuzzaman and Zheng, 2014). Fig. 8 shows the instantaneous time headway distributions for all the vehicles involved in the second campaign based on raw data series. For Peugeot and Mini Cooper, the distribution refers only to ACC and manual driving respectively. From the figures, it becomes clear that all the ACC systems focus on target time headway values, while the distributions without ACC are more uniform and biased to the individual driving behavior. Although initially different headway settings were tried, due to many unexpected cut-ins from other road users and as in the campaign N.1, the time headway setting in the vehicles was set to the shortest value. This is the reason why in some ACC distributions there is a long tail towards higher time headway values. The prevalent value for the ACC controller of the KIA was 1.1s, for the Mitsubishi 1.5s, for the Ford 1.3s, for the Golf 1.4s and for the Peugeot 1.1s. The above observations have direct impact on the traffic flow. From one side, the commercial ACC controllers ensure low speed variations in the longitudinal direction and under stable conditions, but it seems that they cannot efficiently respond to (even small) perturbations that can occur on the road downstream. Car-platoon of human-drivers on the other hand, consume more road capacity on stable conditions but they are efficient in anticipating traffic oscillations. For the sake of completeness the time headway distributions from the first campaign (N.1) are illustrated in Appendix A.

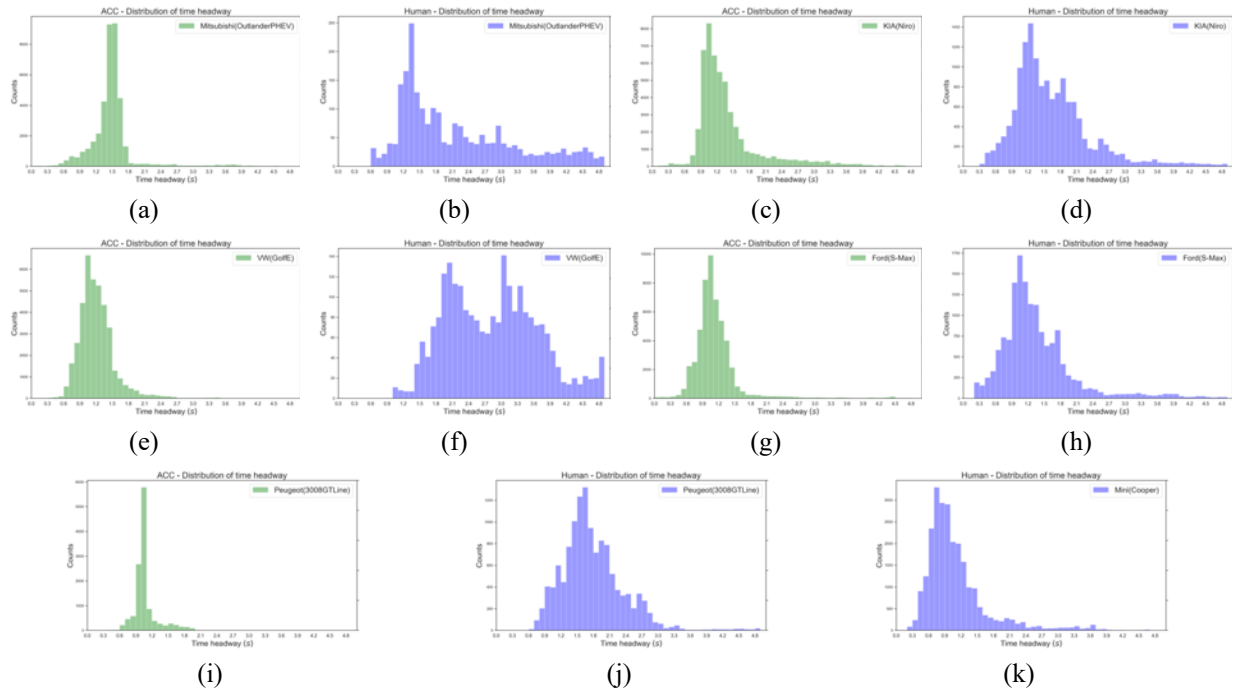

**Figure 8.** Time headway distributions for all the vehicles involved in the campaign N.2.

## 4.2 Response time of the ACC controller

The response time of the ACC controllers is a hot topic for discussion in the literature (Makridis et al., 2019b, 2018; Milanes et al., 2014; Milanés and Shladover, 2014; Shladover et al., 2015). Given the theoretical ability of the ACC systems to retrieve instantly the leader's speed, there is a number of works in the literature that expect instantaneous response time of the controller (Green, 2000; Kesting et al., 2008; Tapani, 2012). This assumption has been adopted by traffic simulation studies. However, empirical tests have shown that systems in commercial vehicles experience significant delays. Makridis et al. (Makridis et al., 2019c) proposed a method for estimation of the response time of an ACC controller based on correlation of two signals, the acceleration of the follower and the speed difference between the leader and the follower. This methodology was used in the present study on the raw data after the application of a moving average filter as discussed above in order to produce more realistic acceleration values for the following vehicle. The goal is to produce some response time estimates for the vehicles involved in the campaigns. We refer the reader to the corresponding publication for more details.

Fig. 9 illustrates the estimated response times for the ACC controllers of two vehicles based on partial trajectories. The top subplots show the two signals that need to be correlated, the delta speed between the two vehicles and the acceleration of the follower. The middle subplot shows the correlation function. The peak corresponds to the correlation coefficient. Values closer to one indicate better confidence on the estimation. In both cases demonstrated here, the response time has been found to be around 2s, which is much larger than the one expected in most of the literature studies. The confidence for both cases is very high and in the third subplot, it is obvious that the two signals become highly correlated for the estimated response times. It is important to note that these estimations refer to normal, non-critical, driving conditions. For safety-related studies the corresponding times of the Automated Emergency Braking ADAS functionality should be considered. However, response time under normal freeway conditions directly impacts traffic and string stability. For the sake of completion, the minimum estimated response time values for the 11 ACC systems are reported in Table IV in Appendix B.

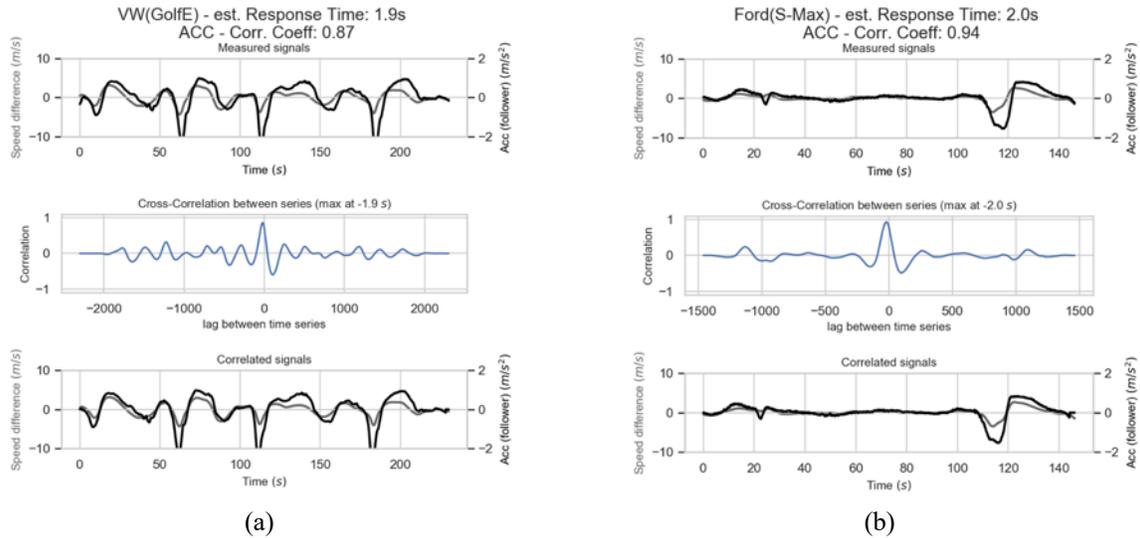

(a) (b)
**Figure 9.** Indicative response time estimations for the ACC controllers of two vehicles from campaign's N.2.

## 4.3 Energy Consumption and Driving Behavior

Another topic where openACC could provide significant insights is the energy consumption of commercial ACC systems in comparison with human drivers. Results from the second experimental campaign have been already published (He et al., 2020a) and are presented here again for completeness.

Two trajectory paths with a five vehicle platoon are studied, one going from the JRC to Vicolungo (Southbound section) and another returning (Northbound section) on the same route with the same vehicles and drivers. (He et al., 2020a). On the southbound section, all vehicles are manually driven, while on the northbound section, the second, third and fourth vehicles are driven by their ACC systems. The impact of the ACC in terms of energy is studied. The results from test vehicles in the middle (second, third and fourth) demonstrate that ACC followers have tractive energy consumption 2.7 – 20.5 % higher than those of human counterparts. The energy consumption was studied also in terms of car-platoon consumption, independently of the vehicle brand and model demonstrating the trend that the vehicles upstream (follower) when driven by ACC, they have higher energy demand. Finally, the correlation in the driving behavior between vehicles in the car-platoon was analyzed. The findings imply that the followers in the automated or mixed traffic flow, relative to those in the manual traffic, generally perform worse in reproducing the driving style of the immediately preceding vehicle. The likely cause for this difference is that the ACC system mainly aims to keep a safe time headway thus shows less flexibility. For more details, we refer the reader to the corresponding publication.

## 4.4 Vehicle dynamics in a controlled environment

The last experimental campaign has been designed to give more a structured and precise dataset using a high-accuracy differential GNSS system for data acquisition. The campaign includes both ACC and Human car-platoon measurements. In all tests, the leading vehicle was driving with the ACC controller in order to ensure as constant as possible speeds and as controlled as possible perturbations. Furthermore, attention was given in ensuring stable conditions for the car-platoon. More specifically, during the experiments, there is radio-based communication between the drivers of the first and the last vehicles in the car-platoon through mobile devices, in order to ensure whenever needed that the car-platoon is in equilibrium state, i.e. all vehicles have the same speed.

In terms of acceleration and deceleration distributions, the data from this campaign are in line with the previous ones, showing that the ACC systems produce low acceleration variations under stable conditions and very sharp ones when the leader performs some perturbation. Also, as expected the difference between raw measurements before and after the application of a moving average filter (similar to the Fig. 4) are much smaller than the on-road campaigns. The corresponding figures are shown for reference in Appendix C.

Some behavioral differences between the ACC systems and human drivers are obvious in Fig. 10. It is interesting to note that under stable car-following conditions there is increased variability around the equilibrium values for position, speed and time headway. Moreover, there seems to be no pattern in the desired time headway values for human drivers, while all ACC systems behave in a very similar way.

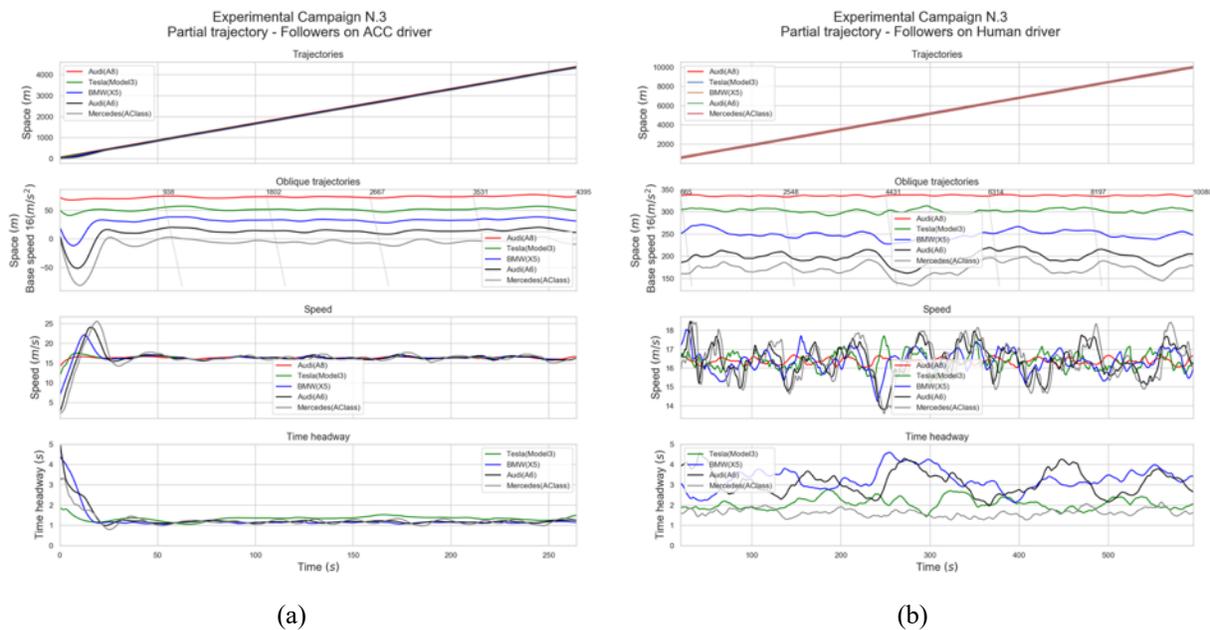

(a) (b)
**Figure 10.** Two snapshots from campaign's N.2 trajectories. In Fig. 8 a) the two followers are driving with ACC on, while in Fig. 8 b) the following vehicle is human driven.

Looking at the time headway distributions across the whole dataset, Fig. 11 illustrates both maximum and minimum headway settings. The peak values found for each vehicles are 1.3s and 2.7s for the Tesla, 1.2s and 2.5s for the BMW, 1.2s and 3.5s for the Audi and 1.2s and 2.5s for the Mercedes. The differences between the time headway distributions of the ACC controller and the human driver are obvious, with the second one being much more uniform. It is worth noting that the ranges of observed time headway values between human and ACC (minimum and maximum boundaries of most observations) are similar, roughly between 1s and 4s.

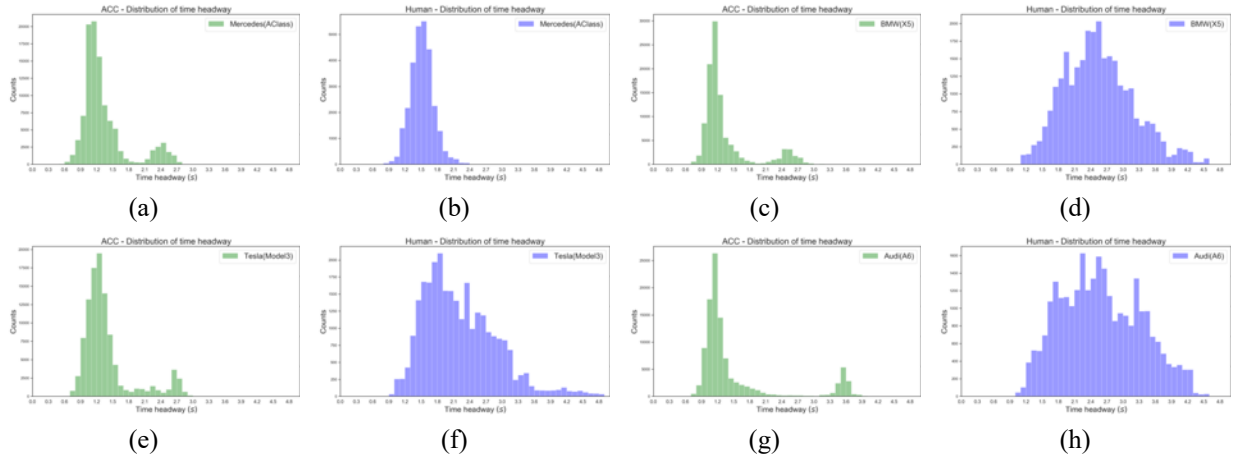

**Figure 11.** Time headway distributions for all the vehicles involved in the campaign N.3.

The insights of this campaign on the indicative response time values for the ACC system of the following vehicles add on the general notion that the commercial systems have high response time values while they are on freeway non-critical conditions. The indicative response times presented here are for the Tesla 1.9s, for the BMW 2.7s, for the Mercedes 2s and for the Audi 2.6s. It has been also observed that the behavior of each controller is not deterministic and varies depending on factors such as road conditions (geometry, altitude), the car-platoon equilibrium speed, the time headway settings, the weather, the magnitude and type of perturbation performed by the leading vehicles and others. However, in all cases the observed response times of the vehicles were by no means instantaneous (i.e. values lower than 0.7s).

## 4.5 String instability and other dimensions

Another interesting topic for researchers is whether commercial ACC systems are string stable or not. A definition of the string stability for a car-platoon assuming that it under stable conditions and the leader performs a perturbation, can be given below: *a car-platoon can be considered string stable when the magnitude of the leading vehicle's perturbation is greater than the magnitude of the following vehicle's perturbation and this holds for all the followers of the car-platoon upstream*.

As it is briefly illustrated in Fig. 10, string instability was observed in the car-platoon. The relation between string instability and time headway settings is currently under study and another topic where the openACC databased could provide some insights. Fig 12 illustrates a partial trajectory from the third experimental campaign where string instability emerges as the leading vehicle fluctuates around the desired speed and initiates a perturbation that propagates upstream. Some findings on the string stability of the commercial ACC systems from the third experimental campaign are available already in the literature (Makridis et al., 2020b). For the sake of completion, an additional illustration of string instability and a comparison with the behavior of human drivers is shown in Fig16 in Appendix D.

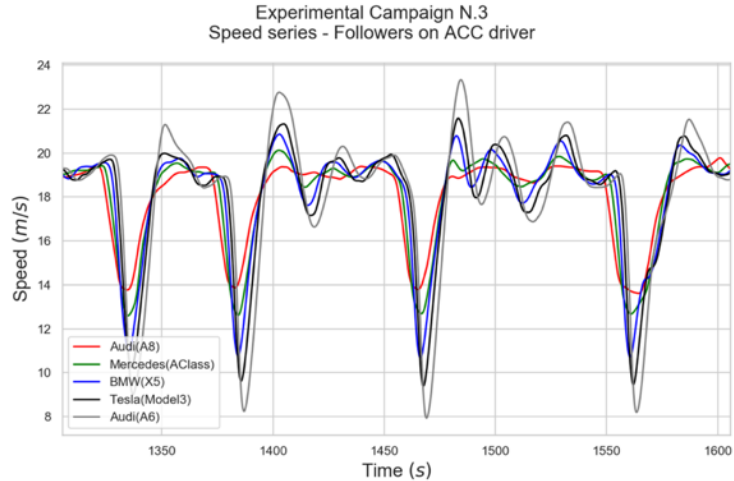

**Figure 12.** Speed series with a car-platoon of 5 vehicles driven by their ACC systems during campaign N.3.

Another research topic of interest where openACC could contribute is a better understanding of the mechanism of traffic hysteresis and traffic oscillations from the driver behavior perspective as described in the literature (Laval and Leclercq, 2010; Saifuzzaman et al., 2017). In the third campaign dataset, the acceleration, deceleration and stable states are separated by experimental design. This characteristic facilitates the study of traffic hysteresis and its magnitude as demonstrated in Figure 13a. Here, the three main phases (i.e. deceleration, acceleration, equilibrium) are colored differently, showing the differences in the density during the non-stable conditions as well as the different behavior of maximum and minimum headway settings of the controllers. For the minimum headway setting, during the acceleration phase, the density is much smaller than during the deceleration phase, for similar speeds, so the produced flow is decreased. On the other hand, there is no such phenomenon observable for the maximum headway setting. However, the flow is already smaller than even the decreased flow produced by the minimum headway platoon during the acceleration phase.

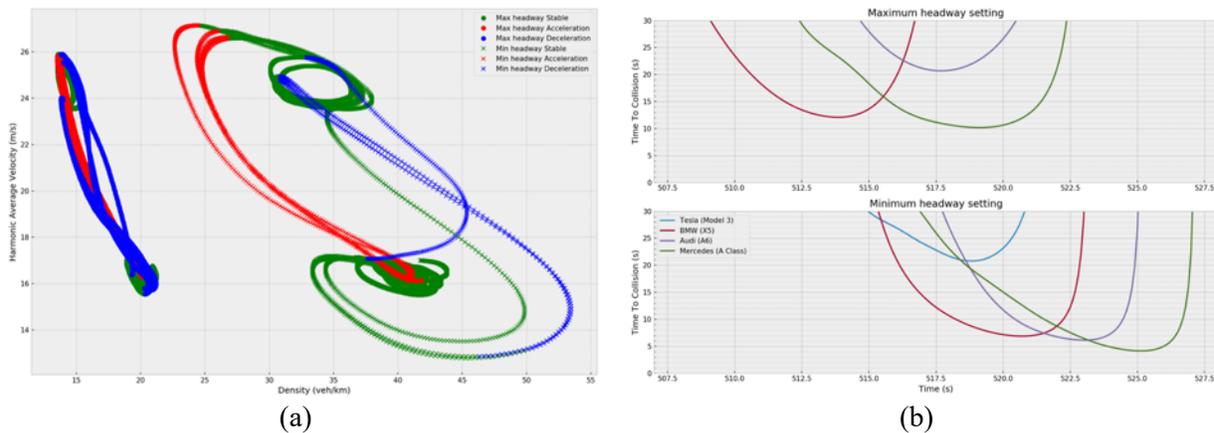

(a)                                                                                       (b)

**Figure 13.** Demonstration of traffic hysteresis based on the data from the campaign N.3.

Analysis regarding traffic safety can be also performed looking at common Safety Surrogate Metrics such as Time-to-Collision (TTC) as shown in Figure 13b. Here the TTC instantaneous values are shown for two different headway settings in the vehicles of the third campaign. The difference in the instantaneous TTC values for the vehicles with the maximum and minimum headway settings is clear. It is worth noticing that with the minimum headway setting the minimum TTC decreases moving backward in the platoon of vehicles. This is clearly related to the string instability and the need for vehicles further upstream to decelerate stronger than the vehicles upstream. Looking at the particular case presented in Figure 13b the last vehicle in the platoon has minimum TTC of 4s corresponding to a no-risk situation (literature suggests a crisp threshold of 1.5s to identify risky situations). However, if the platoon was longer it could be expected that vehicles further backward would have achieved high collision risk even if the deceleration of the leader was extremely mild (as mentioned in Section 3 accelerations and decelerations are applied changing the target speed in the ACC of the leader vehicle). As the number of ACC equipped vehicles increases, the probability to have on our motorways platoons of ACC-enabled vehicles will increase with possibly consequent negative effect both on traffic stability and safety.

# 5 Discussion and Conclusions

This work summarizes the main features of the openACC, an open-access database of different car-following experiments involving a total of 11 vehicles equipped with state-of-the-art commercial ACC systems. As more test-campaigns will be carried out and the related data extracted, openACC will evolve accordingly. The activity is carried out in the framework of the openData policy of the European Commission Joint Research Centre with the objective to engage the whole scientific community towards a better understanding of the properties of ACC vehicles in view of anticipating their possible impacts on traffic flow and prevent possible problems connected to their widespread.

This open science dataset creates unique opportunities for researchers in order to observe commercially available automation controllers under real world conditions. Preliminary results from each dataset have been presented, along with few concrete examples on the topics where researchers can work on, some of which our outside the authors' research activities. The database consists of a series of experimental campaigns focused on traffic related ACC properties based on car-following tests. It involves 11 ACC-equipped vehicle models of different brands that are currently available on the market and 6 more vehicles used as leaders in the tests. Two campaigns refer to tests on public roads and another one in a proving ground facility. The data acquisition was performed with two different systems and after post processing the frequency of the measurements is 10hz for all the cases. The main properties that have been discussed here are the acceleration and deceleration operational domain of commercial ACC systems, prevalent time headway values, estimated observable response times for the controllers and sting instability problems that arise from small perturbations downstream in an ACC car platoon.

A more structured and systematic study of the dataset should be performed in order to validate and confirm the preliminary findings discussed in this work. However, the findings are convincing enough to the following main conclusions.

For the commercially available ACC systems:

- They produce smooth acceleration and deceleration values under stable conditions.
- Speed perturbations of the leader in a car-platoon show that they are string unstable, especially for small time headway settings.
- Their response times under non-critical driving conditions are certainly non-negligible and in any case comparable of not higher than those of attentive human drivers.
- Their minimum and maximum headway values are around 1s and 2.5s.
- Headway settings should be standardized, as high penetration of such systems will heavily impact road transport networks.

- Their driving behavior is more homogenous than human drivers.

For the human drivers:

- There is high variance in the behavior of different individuals in terms of time headway, acceleration and deceleration strategy, and response time to perturbations upstream.
- They can anticipate perturbations upstream, adapting their driving strategy accordingly and absorbing traffic oscillations.
- The variability in the driving behavior of individuals creates free-flow pockets even under car-following conditions.
- Microsimulation models should take into consideration the variability in the driving behaviors and the vehicle dynamics.

Future plans involve data acquisition campaigns with longer car-platoons in order to shed more light on the string instability of these ADAS systems, data acquisition of Automated Emergency Braking measurements to observed the impact of automation under critical conditions, study of other ADAS systems such as lane keeping, lane changing, chauffeur etc. and deeper investigation on the impact of automated vehicle functionalities on traffic and energy demand. On the post-processing side, a more in-depth study is needed to identify a filtering technique able to improve the quality of the existing data in order to better represent the car-platoon consistency. Finally, since most traffic models use parameters with physical meaning such as reaction time or maximum acceleration, empirical campaigns such as those described in this paper, should be used to find the correlation between model inputs and observations. More robust microsimulation models should be developed considering in an explicit way the variability in the human driving behaviors and the impact of realistic vehicle acceleration dynamics on traffic flow and energy demand.

# ACKNOWLEDGMENTS

The views expressed here are purely those of the authors and may not, under any circumstances, be regarded as an official position of the European Commission. The authors are grateful to Eleftherios-Nektarios Grylonakis and Vincenzo Arcidiacono who contributed to the data processing, as well as to Georgios Fontaras, Fabrizio Minarini, Norbert Brinkhoff-Button, Gianmarco Baldini, Daniele Borio, Alessandro Tansini, Melania Susi, Lorenzo Maineri, Yinglong He, Akos Kriston, and Fabrizio Re for their support during the experimental campaigns. Finally, the authors would like to thank all the stuff in AstaZero proving ground in Sweden for their support during the experiments.

Makridis et al., 2020                                                                                                         23Makridis, M., Fontaras, G., Ciuffo, B., Mattas, K., 2019a. MFC Free-Flow Model: Introducing Vehicle Dynamics in Microsimulation. Transp. Res. Rec. 2673, 762–777. https://doi.org/10.1177/0361198119838515

Makridis, M., Mattas, K., Borio, D., Ciuffo, B., 2019b. Estimating empirically the response time of commercially available ACC controllers under urban and freeway conditions, in: 2019 6th International Conference on Models and Technologies for Intelligent Transportation Systems (MT-ITS). Presented at the 2019 6th International Conference on Models and Technologies for Intelligent Transportation Systems (MT-ITS), pp. 1–7. https://doi.org/10.1109/MTITS.2019.8883341

Makridis, M., Mattas, K., Borio, D., Giuliani, R., Ciuffo, B., 2018. Estimating reaction time in Adaptive Cruise Control System, in: 2018 IEEE Intelligent Vehicles Symposium (IV). Presented at the 2018 IEEE Intelligent Vehicles Symposium (IV), pp. 1312–1317. https://doi.org/10.1109/IVS.2018.8500490

Makridis, M., Mattas, K., Ciuffo, B., 2019c. Response Time and Time Headway of an Adaptive Cruise Control. An Empirical Characterization and Potential Impacts on Road Capacity. IEEE Trans. Intell. Transp. Syst. 1–10. https://doi.org/10.1109/TITS.2019.2948646

Makridis, M., Mattas, K., Ciuffo, B., Fontaras, G., 2020a. The impact of automation and connectivity on traffic flow and CO2 emissions. A detailed microsimulation study.

Makridis, M., Mattas, K., Ciuffo, B., Re, F., Kriston, A., Minarini, F., Rognelund, G., 2020b. Empirical Study on the Properties of Adaptive Cruise Control Systems and Their Impact on Traffic Flow and String Stability. Transp. Res. Rec. 0361198120911047. https://doi.org/10.1177/0361198120911047

Mattas, K., Makridis, M., Hallac, P., Raposo, M.A., Thiel, C., Toledo, T., Ciuffo, B., 2018. Simulating deployment of connectivity and automation on the Antwerp ring road. IET Intell. Transp. Syst. 12, 1036–1044. https://doi.org/10.1049/iet-its.2018.5287

Milanés, V., Shladover, S.E., 2014. Modeling cooperative and autonomous adaptive cruise control dynamic responses using experimental data. Transp. Res. Part C Emerg. Technol. 48, 285–300. https://doi.org/10.1016/j.trc.2014.09.001

Milanes, V., Shladover, S.E., Spring, J., Nowakowski, C., Kawazoe, H., Nakamura, M., 2014. Cooperative Adaptive Cruise Control in Real Traffic Situations. IEEE Trans. Intell. Transp. Syst. 15, 296–305. https://doi.org/10.1109/TITS.2013.2278494

Munoz, J.C., Daganzo, C.F., 2002. Fingerprinting Traffic From Static Freeway Sensors.

Ngoduy, D., Lee, S., Treiber, M., Keyvan-Ekbatani, M., Vu, H.L., 2019. Langevin method for a continuous stochastic car-following model and its stability conditions. Transp. Res. Part C Emerg. Technol. 105, 599–610. https://doi.org/10.1016/j.trc.2019.06.005

NGSIM [WWW Document], 2006. URL https://ops.fhwa.dot.gov/trafficanalysistools/ngsim.htm (accessed 12.11.19).

Olia, A., Razavi, S., Abdulhai, B., Abdelgawad, H., 2018. Traffic capacity implications of automated vehicles mixed with regular vehicles. J. Intell. Transp. Syst. 22, 244–262. https://doi.org/10.1080/15472450.2017.1404680

SAE International, 2018. J3016B: Taxonomy and Definitions for Terms Related to Driving Automation Systems for On-Road Motor Vehicles - SAE International [WWW Document]. URL https://www.sae.org/standards/content/j3016_201806/ (accessed 4.9.20).

Saifuzzaman, M., Zheng, Z., 2014. Incorporating human-factors in car-following models: A review of recent developments and research needs. Transp. Res. Part C Emerg. Technol. 48, 379–403. https://doi.org/10.1016/j.trc.2014.09.008

Saifuzzaman, M., Zheng, Z., Haque, Md.M., Washington, S., 2017. Understanding the mechanism of traffic hysteresis and traffic oscillations through the change in task difficulty level. Transp. Res. Part B Methodol. 105, 523–538. https://doi.org/10.1016/j.trb.2017.09.023

# Appendix

## A. Time headway distributions from the campaign N.1

The time headway distributions for the human driver and the ACC system of the two vehicles driving as followers in the car-platoon during the first campaign are shown in Figure 14.

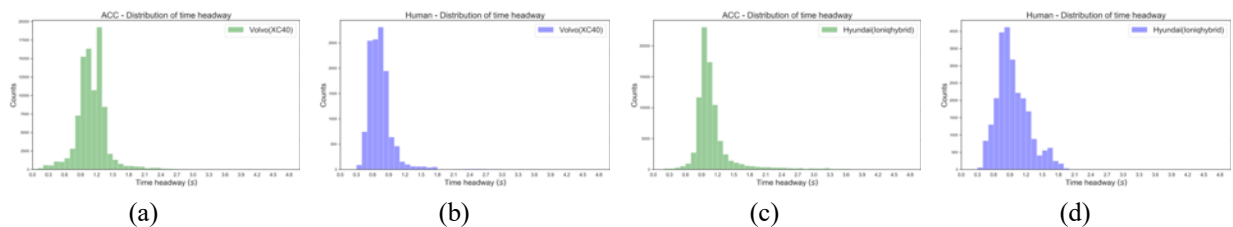

(a)                    (b)                    (c)                    (d)

**Figure 14.** Time headway distributions for all the vehicles involved in the campaign N.1.

## B. Minimum response time estimates

Indicative response time values for the 11 ACC systems under test are presented as a reference in Table IV. It should be mentioned that the controllers' functionality is by no means deterministic and consequently the values reported here can be perceived only as estimation that show an order of magnitude for the response time of the controllers. The values presented here are the minimum values estimated under freeway non-critical conditions for the controllers. As shown in the table below the ACC response times are very similar to human response time and by no means instantaneous.

**Table IV** Minimum response time estimates for the 11 ACC systems under study.

|  | Estimated ACC Response Time | Campaign |
|---|---|---|
| Volvo (XC40) | 2s | N.1 |
| Hyundai (Ioniq hybrid) | 1.6s | N.1 |
| KIA (Niro) | 1.8s | N.2 |
| Mitsubishi (Outlander PHEV) | 1.9s | N.2 |
| Peugeot (5008 GT Line) | 1.7 | N.2 |
| VW (Golf E) | 1.9s | N.2 |
| Ford (S-Max) | 1.7s | N.2 |
| Tesla (Model 3) | 1.7s | N.3 |
| BMW (X5) | 2.6s | N.3 |
| Mercedes (A Class) | 1.9s | N.3 |
| Audi (A6) | 1.9s | N.3 |

## C. Acceleration distributions with differential GNSS

Figure 15 illustrates the acceleration distributions for the campaign conducted in a test track based on raw data measurements, before and after the application of a moving average filter. In contrast to on-road tests, data post-filtering is not necessary for the third campaign in order to produce accelerations within realistic boundaries.

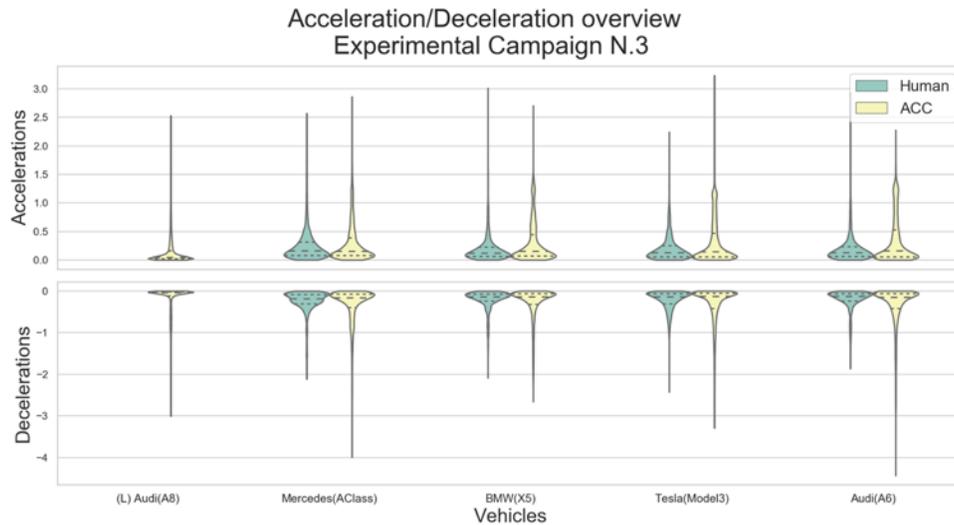

(a)

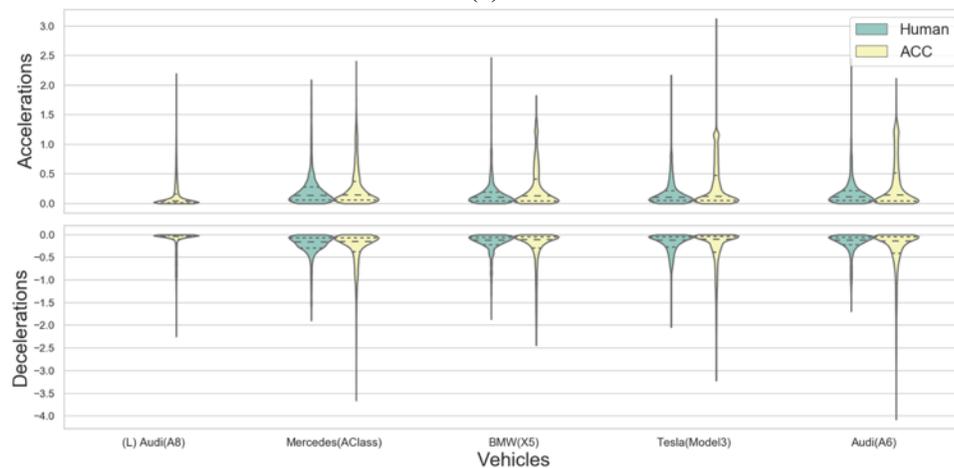

(b)

**Figure 15.** Acceleration and deceleration distributions per vehicle and driving mode. Prefix "(L)" before the name of a vehicle signifies that this vehicle was used as a leader of the car-platoon. Fig. 14a) depicts the raw measurements, while Fig. 14b) illustrates the same data after the application of a moving average filter.

## D. String instability of ACC systems

Figure 16 provides four examples with speed series from the second and third campaign. Fig. 16a shows a five-vehicle platoon from campaign N.2 where all the followers (4 vehicles) were driving with ACC on. String instability is obvious for the car-platoon. This is not obvious in Fig. 16b when all the vehicles are driven by human drivers. A much more obvious example is shown in the figures 16c and 16d that correspond to partial trajectories from campaign N.3. Here, the differences between ACC systems and Human drivers are more apparent. However, it should be mentioned that since the tests were conducted in a controlled environment, by professional drivers, the car-platoon with human drivers might become more unstable under real-world situations.

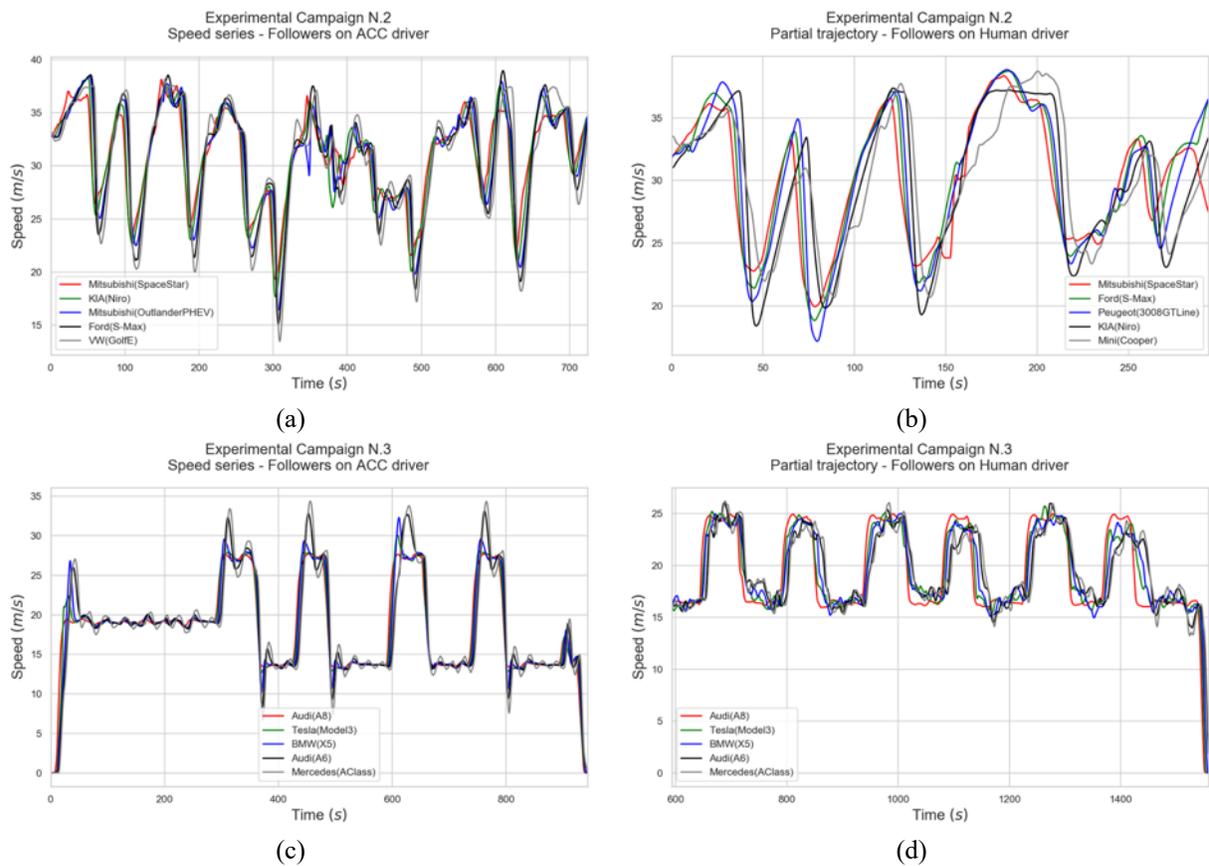

**Figure 16.** The behavior of ACC systems and Human drivers regarding car-platoon instability for the campaigns N.2 and N.3.

Makridis et al., 2020                                                                                                                  27

# E. Indicative template of the information file

Each folder inside the openACC database contains a file with the specifications of the vehicle as well as a short description of the experimental campaign. Below a sample is presented on how the information file is structured.

*Description of the experiment: A main description of the campaign, and objectives. The number of trips included.*

*Date: DD/MM/YYYY.*

*Vehicle specs: Name of the vehicles and vehicles' specifications.*

*Number of vehicles: Number of vehicles involved in each campaign.*

*Equipment: Data acquisition equipment.*

*Data processing: Technique used for the 10 Hz frequency achievement.*

*Trip comments: Some comments on the trips on the data of the trips.*

*Columns info: Description of the columns included in the csv files regarding the respective experimental campaign.*